\documentstyle[11pt]{article}
\begin{document}
\begin{titlepage}

\vspace{0.5cm}

\centerline{\Large \bf  Variational Estimation of the Wave Function at Origin}
\centerline{\Large \bf  for Heavy Quarkonium\footnote{This work was
partly supported by the National
Natural Science Foundation of China (NSFC) and Istituto Nazionaale di Fisica
Nuclear of Italy (INFN) }}
\vspace{1cm}
\centerline{ Yi-Bing Ding$^{a,b,c,d}$,
             Xue-Qian Li$^{e,a,c}$, Peng-Nian Shen$^{f,a,c}$}
\vspace{0.5cm}
{\small
{
\flushleft{\bf  $~~a.$ China Center of Advanced Science and Technology
 (World Laboratory),}
\flushleft{\bf  $~~~$P.O.Box 8730, Beijing 100080, China}
\vspace{8pt}
\flushleft{\bf  $~~b.$ Graduate School, USTC at Beijing,
Academia Sinica}
\flushleft{\bf  $~~~$P.O. Box 3908, Beijing 100039, China}
\vspace{8pt}
\flushleft{\bf  $~~c.$ Institute of Theoretical Physics, Academia Sinica,
 P.O.Box 2735,}
\flushleft{\bf  $~~~$Beijing 100080, China}
\vspace{8pt}
\flushleft{\bf  $~~d.$ Department of Physics, University of Milan, INFN}
\flushleft{\bf  $~~~$Via Celoria 16, 20133 Milan, Italy}
\vspace{8pt}
\flushleft{\bf  $~~e.$ Department of Physics, Nankai University, Tianjin
300071, China}
\vspace{8pt}
\flushleft{\bf  $~~f.$ Institute of High Energy Physics, Academia Sinica,
 P.O.Box 918(4),}
\flushleft{\bf  $~~~$Beijing 100039, China}
}}

\vspace{1cm}

\begin{minipage}{11cm}

\centerline{\bf Abstract}

  {\rm

\noindent

The wave function at the origin (WFO) is an important quantity in studying
many physical problems concerning heavy quarkonia. However, when one used
the variational method with fewer parameters, in general, the deviation of
resultant WFO from the "accurate" solution was not well estimated. In
this paper, we discuss this issue by employing several potential forms
and trial wave functions in detail and study the relation between  WFO
and the reduced mass.

}

\vspace{0.5cm}

{\bf PACS number(s): 12.39.Jh, 12.39.Pn, 03.65.Ge }

\end{minipage}
\end{titlepage}

\baselineskip 18pt

\noindent {\bf I. Introduction}

\vspace{0.5cm}

Recently, the wave function at the origin for the S-wave bound state of a
heavy quark-antiquark system once again attracts physicists'
attentions$^{\cite{cd,eq}}$. This is because that it is not only a very
important quantity for calculating spin state hyperfine splitting, but
also crucial to evaluating the production and decay amplitude of the
heavy quarkonium. Within the context of the non-relativistic potential
model, Refs. \cite{cd} and \cite{eq} demonstrated the numerical results
of WFO of the S-wave $c \bar c$ ,  $b \bar c$  and $b \bar b$ systems
and compared with those obtained in various "successful" potential models.

\vspace{0.3cm}

As well known, except the Coulomb and the harmonic oscillator potentials,
there are few potentials which bound state problems can be analytically
solved. For solving these non-analytically soluble bound state problems,
one has to use approximations. Numerically solving Schr\"{o}dinger equation
is the most powerful method which can reach most required accuracy.
But the numerical method has some defects, for instance, it cannot give
analytical expressions for further discussion. Moreover, all numerical method
for the central potential  is only available for the $v(r) $ which has the
singularity less than $\frac {1}{r^2}$ when $r$ approximate to $0$,
therefore, it definitely fails as $\frac{1}{r^3}$ exist in the potential.
It is unfortunately, the case is encountered usually in calculating the
fine-splitting of the P state.

\vspace{0.3cm}

The perturbative method is another approximation method which has most
extensively been used. However, the practical applicability of the
perturbative expansion in many cases is limited due to divergence, and the
ranges of perturbation parameter values are usually restricted by the
convergence requirements. Moreover, in the perturbation method, the
treatment  for wave functions is much more difficult than that for energy
eigenvalues. As it was declared by MacClary and Byers \cite{mb}
that it is not simple to obtain the wave function correct to the
order of $v^2/c^2$, because the perturbative correction to the wave
function should be given by an infinite sum over all states of the
nonrelativistic Hamiltonian. As an example, in the positronium case,
this sum  diverges because of the divergence  of the potential at the
origin. Besides, for the singular $\frac{1}{r^3}$ potential, the perturbative
method is the one used most frequently yet. The splitting of the energy level
is obtained by calculating an integral. However, as mentioned in
Ref.\cite{lan}, such a singular potential would lead to an exotic result
which does not correspond to the real physics.  Therefore, the meaning of the
result given by the perturbative calculation is not evident actually.

\vspace{0.3cm}

Of course, as a physical state which can be measured in the experiment, there
should be a theoretically derived solution to match, as long as the employed
model is correct. The problem is how to get it. For instance, the energy
levels of the triplet P states of the $c \bar{c}$ system must have definite
measured values. In the theoretical calculation, due to the non-soluble
nonperturbative QCD effect, one has to rely on specific models.
The non-relativistic quark potential model is one of them.
Under the tree diagram approximation and the non-relativistic approximation,
a potential term of $r^{-3}$ appears.  This is simply because that even in the
perturbative framework, one does not collect (actuarially is not able to
do so) all the high order diagrams in the perturbative expansion and all
the high order terms in the non-relativistic reduction. Then the question
is how to catch the major character of the real physics.

\vspace{0.3cm}

There were lots of conscientious attempts have been made. It is noteworthy
that in Gupta's papers$^{\cite {grp,gjr}}$, the non-relativistic reduction
was performed with respect to the power of $\vec{p}^2/E^2$ so that the
singular $r^{-3}$ potential term can be avoided. Our recent study indicates
that the variational method can give some interesting sight to the singular
$\frac{1}{r^3}$ problem.
\vspace{0.3cm}

As well known that the variational methods a widely employed approach.
In principle, by using this method one can leave the above mentioned problem
alone and
get the major content of the real physics. We will discuss this issue in
our successive paper \cite{dls}. In spite of this, the variational
method has more advantages. It can give an analytical expression of
the wave function. In particular, if there is only a single parameter
in the trial wave function, the resultant wave function has a simple
form. Then it is very convenient in the practical application and
physical discussion.

\vspace{0.3cm}

The variational method has extensively been used and seems to be successful
in many aspects. Although most of works paid their attentions on seeking out
accurate energy eigenvalues and seldom discussed the wave function,
in particular WFO,  Ref. \cite{sch} simply discussed the wave function by
employing the multi-Gaussian trial wave function.

\vspace{0.3cm}

In many simple cases, the Gaussian-type or the exponential-type functions
with a single-parameter were taken as the trial wave function to discuss
the ground or excited states. In general, the accuracy of the resultant
energy eigenvalues is satisfactory. However, by looking at the wave
function, one would find that although the resultant wave function
sometimes may not deviate much from the "accurate" solution (usually it can
be obtained by solving  the Schr\"{o}dinger equation numerically) in the
long-range part, but this deviation turns larger and larger when
$r\rightarrow 0$.

\vspace{0.3cm}

In this work, we would study this problem in detail with various potentials
which were used frequently in studying heavy quarkonia and put more
emphasis on estimating WFO. The adopted trial wave functions were used
in literatures except the last one.

\vspace{0.3cm}

The paper is organized as following. After the introduction, the variational
method by using the single$-$parameter trial wave function is discussed
in Sec.II. In Sec.III, variational calculation by employing multi-parameter
trial wave functions is presented. The variational study on the 2S state of
the $c\bar c$ system is further shown in Sec.IV, and in Sec.V, the relation
between WFO and the reduced mass for various trial wave functions is
investigated. In the last section, the discussion and conclusion are given.

\vspace{0.5cm}

\noindent {\bf II. Variational method with  single$-$parameter
trial wave function}

\vspace{0.5cm}

There are many potential models which can fit the experimental spectra of
the heavy quarkonia with certain accuracy. In the rest of the paper, we
call them as "successful" potential models.

\vspace{0.3cm}

Within the framework of the non-relativistic potential model, the S-state
wave function $\psi(r)$ of the heavy quarkonium satisfies the Schr\"{o}dinger
equation
\begin{eqnarray}
H\psi(r)=-\frac{1}{2\mu}\Delta \psi(r)+V(r)\psi(r)=E\psi(r)
\end{eqnarray}
where $H$ is the Hamiltonian of the quarkonium, $V(r)$  denotes the central
potential between quark and antiquark, $E$ represents the
energy eigenvalue, and $\mu$ is the reduced mass.
In general, one can numerically solve the Schr\"{o}dinger equation
to obtain $E$ and $\psi(r)$ simultaneously, and calculates $\psi(0)$ (WFO)
in terms of the average value of $\frac{dV}{dr}$ \cite{qr}, i.e.
\begin{eqnarray}
|\psi(0)|^2=\frac{\mu}{2 \pi} <\frac{dV}{dr}>
\end{eqnarray}
The values of the squared WFO for various "successful" potential models
were listed in Refs. \cite{cd,eq}. (note: $|R(0)|^2$ is equal to $4 \pi
 |\psi(0)|^2$ in the Table 1 of Ref.\cite{cd}. ).

\vspace{0.3cm}

To solve Eq.(1) by using the variational method, one needs to choose a
suitable trial wave function $\psi(r; c)$ with $N$ independent parameters
$\{c\}=\{c_1,c_2,\cdot\cdot\cdot,c_N\}$ first and then to seek out
a set of parameters $\{c_0\}=\{c_{i0}, i=1,2,\cdot\cdot\cdot,N\}$ which
minimizes the expectation value of Hamiltonian, namely
\begin{eqnarray}
E(c)=<H>=\frac{<\psi(c)|H|\psi(c)>}{<\psi(c)|\psi(c)>}.
\end{eqnarray}
The minimum value $E(c_0)$ gives an upper limit of the ground state energy.
In the use of the variational method, one wishes to obtain  $E(c_0)$ which
is as close as possible to the "accurate" solution with the minimum number
of parameters.

\vspace{0.3cm}

In this work, what we concern is how close to the "accurate" solution
the resultant wave function can be, when $E(c_0)$ is satisfactorily
close to the "accurate" energy eigenvalue. In particular, we would try
to find the accuracy of WFO for various trial wave functions, which
obviously affects the application of the variational method.

\vspace{0.3cm}

In this paper, we choose three
most popular and "successful" models listed in Refs.\cite{cd,eq}, so that
the conclusion could be more general. These models are \footnote{Indeed, k
in Eqs.(5) and (6) should have proper dimensions, but in our work, it is
not important because they would be attributed into the normalization
factors.}:

(1). Cornell potential\cite{egkl}:
\begin{eqnarray}
V(r)=-\frac{4}{3} \frac{\alpha_s}{r}+ k r,
\end{eqnarray}
with $\alpha_s=0.39$,  $k=1/2.34^2(GeV)^2$ and the mass of $c$
quark $m_c=1.84 GeV$.

(2). Martin potential\cite{mar}:
\begin{eqnarray}
V(r)=kr^{0.1},
\end{eqnarray}
with $k=6.898$ and $m_c=1.8 GeV$.

(3). Logarithmic potential\cite{qr1}:
\begin{eqnarray}
V(r)=k log(r),
\end{eqnarray}
with $k=0.733$ and $m_c=1.5 GeV $.

\vspace{0.3cm}

In this section, we choose the simplest trial wave function in which
there is only one variational parameter to study the $1S$ state of $c\bar{c}$.
The general form of such trial wave function is written as

\begin{eqnarray}
     \psi_{trial}(r)=N~ e^{-a~r^b}
\end{eqnarray}
where

\begin{eqnarray}
    N=[\frac{b~(2a)^{\frac{3}{b}}}{4 \pi \Gamma(\frac{3}{b})}]^{\frac{1}{2}}.
\end{eqnarray}
is the normalization constant, $a$ denotes the variational parameter
which will be fixed by minimizing the expectation value of
Hamiltonian and $ b $ is the model parameter which determines the
type of the trial wave function. In practice, we select following four
trial wave functions:

(1). $b=1$, namely $N~e^{-a~r}$ (hydrogen wave function or exponential wave
function). It is the solution of the Coulomb potential model.

(2). $b=2$, namely $N~e^{-a~r^2}$ (harmonic oscillator wave function or
a Gaussian wave function).

(3). $b=\frac{3}{2}$, namely  $N~e^{-a~r^{3/2}}$. This function was used
by Gupta \cite{grp}.

(4). $b=\frac{4}{3}$, namely $N~e^{-a~r^{4/3}}$. This is a newly proposed
trial wave function, and we will pay more attention on it.

\vspace{0.3cm}

In order to fully understand the accuracies of the variational results
in  three different potential cases, we calculate four quantities by using
four different trial wave functions and the corresponding "accurate" results
by  solving the Schr\"{o}dinger equation numerically as well.
These quantities are:

\vspace{0.3cm}

(1). the energy eigenvalue $E$ (note: we do not try to fit the
experimental spectrum here, because it is not the aim of this work).

(2). average radius $<r>$.

(3). average value of the inversed  radius  $<\frac{1}{r}>$.

(4) the squared WFO $|\psi (0)|^2$.

\vspace{0.3cm}

For each calculated quantity $q$ we give a relative deviation
$\delta q$, which is defined as
$$\delta q=\frac{(q_{var} -q_{true})}{q_{true}},$$
where $q_{var}$ is the variational result and $q_{true}$ the "accurate"
value.

\vspace{0.3cm}

In the case of the Cornell potential (4),  kinetic energy $<T>$ and
potential energy $<V>$ are:
\begin{eqnarray}
 <T> &=& \frac {(2a)^{\frac{2}{b}}b^2 \Gamma (2+\frac{1}{b})}{8 \mu
 \Gamma(\frac{3}{b})}, \\
<V> &=& \frac{-4(2a)^{\frac{2}{b}} \alpha_s \Gamma(\frac{2}{b})+
3 k \Gamma(\frac{4}{b})}{3 (2a)^{\frac{1}{b}} \Gamma(\frac{3}{b})}.
\end{eqnarray}

Then we can obtain the expectation value of Hamiltonian and consequently
an algebraic equation, which is used to determine $a$,
$$ [3 b^2 \Gamma(2+\frac{1}{b})] x^3-
     [16 \alpha_s \mu \Gamma(\frac{2}{b})] x^2
            -12 k \mu \Gamma(\frac{4}{b})=0,  $$
where  $x=(2a)^{\frac{1}{b}}$.
It is very easy to solve  this equation. If we rewrite it in the
following form:
\begin{eqnarray}
A_3 x^3+A_2 x^2+A_0=0 ,
\end{eqnarray}
the real solution of $x$ can be expressed as:
\begin{eqnarray}
  x_{real}=-\frac{A_2}{3A_3}+\frac{2^{\frac{1}{3}} A_2^2}{3A_3 B}+
           \frac{B}{3~~ 2^{\frac{1}{3}} A_3},
\end{eqnarray}
where $B=(B_0+\sqrt{-4 A_2^6 +B_0^2})^{\frac{1}{3}}$ and
$B_0\equiv -2A_2^3-27A_0A_3^2$.

\vspace{0.3cm}

In the case of Martin potential (5), the potential energy reads:
\begin{eqnarray}
<V>=\frac {k\Gamma(\frac{3.1}{b})}{(2a)^{\frac{0.1}{b}}\Gamma(\frac{3}{b})}.
\end{eqnarray}
Therefore, the equation for determining $a$ is quite simple. The solution is
\begin{eqnarray}
a=\frac{1}{2}[\frac{0.4~ \mu~  k~\Gamma(\frac{3.1}{b})}
          {b^2\Gamma(2+\frac{1}{b})}]^{\frac{b}{2.1}}.
\end{eqnarray}

\vspace{0.3cm}

In the case of logarithmic potential (6), by
the similar procedure we  obtain
\begin{eqnarray}
a=\frac{1}{2}[\frac{4 \mu~k~\Gamma(\frac{3}{b})}
                   {b^2~\Gamma(2+\frac{1}{b})}]^{\frac{b}{2}}.
\end{eqnarray}

\vspace{0.3cm}

All the numerical results are listed in  Tables 1.1 to 1.3. In order to make
comparison, we write the corresponding "accurate" results in the table
captions and list relative deviations in the corresponding tables.

\vspace{0.5cm}
\centerline{ Table 1.1 }
\vspace{0.3cm}
{\footnotesize
The variational results with a single parameter trial function in
the Cornell potential case. The values listed in parentheses
 are relative deviations. The
"accurate" results are: $E(1S)=0.257526\; GeV$, $<r>=1.7073\; GeV^{-1}$,
$<\frac{1}{r}>=0.80848\; GeV$ and $|\psi(0)|^2=0.116054\; GeV^3$.  }

\begin{footnotesize}
\begin{center}
\begin{tabular}{|c|c|c|c|c|}
\hline
  $b$ &  $ E$      & $ <r>$   &     $ <\frac{1}{r}>$ &   $|\psi(0)|^2$\\
\hline
1& 0.272795~( 0.059) &  1.7908~ ( 0.049) &  0.8376~ (  0.036) & 0.197052~( 0.61)\\
\hline

2& 0.280039~( 0.087) &  1.7234~ ( 0.009) &  0.7388~ ( -0.098) & 0.050403~( -0.57) \\
\hline

$\frac{3}{2}$& 0.259785~( 0.0088)&  1.6989~ (0.0049)& 0.7908~( -0.022)& 0.082911( -0.29)\\
\hline

$\frac{4}{3}$& 0.257809~ (0.0011)& 1.7083~( 0.0006)& 0.8083~ (-0.0002)& 0.103334( -0.11)\\
\hline
\end{tabular}
\end{center}
\end{footnotesize}
 
\vspace{0.3cm}
\centerline{ Table 1.2 }
\vspace{0.3cm}

{\footnotesize

The variational results with a single parameter trial function in the Martin
potential case. The values listed in parentheses are relative deviations.
The "accurate" solutions are: $E(1S)=7.5605\; GeV$,
$<r>=1.72332\; GeV^{-1}$,
$<\frac{1}{r}>=0.782946\; GeV$, $|\psi(0)|^2=0.0778779\; GeV^3$.    }

\begin{footnotesize}
\begin{center}
\begin{tabular}{|c|c|c|c|c|}
\hline

$ b$ &  $ E$   &  $ <r>$ & $ <\frac{1}{r}>$  &     $|\psi(0)|^2$\\
\hline
           
1 & 7.5871~( 0.0035)& 1.8606~(0.079) & 0.8064~( 0.030) & 0.166933~( 1.14) \\
\hline

2    & 7.57509~( 0.0019)&1.7151~(-0.0048)& 0.7423~(-0.052)& 0.051445~(-0.34)\\
\hline

$\frac{3}{2}$ &7.5607~($3*10^{-5}$)&1.7194~(-0.0023)& 0.7814~(-0.0020)& 0.079984(-0.027)\\
\hline
$\frac{4}{3}$ &7.5622~(0.0002)&1.7415~(0.011)& 0.7928~(0.013)&0.097596~(-0.25)\\
\hline
\end{tabular}
\end{center}
\end{footnotesize}
\vspace{0.3cm}
\centerline{ Table 1.3 }
\vspace{0.3cm}

{\footnotesize

The variational results with a single parameter trial function in the
logarithmic potential case. The values listed in parentheses are  relative
deviations. The "accurate" solutions are: $E(1S)=0.730733\; GeV$,
$<r>=1.87535\; GeV^{-1}$,
$<\frac{1}{r}>=0.723538\; GeV$,
$|\psi(0)|^2=0.0633063\; GeV^3$.   }

\begin{footnotesize}
\begin{center}
\begin{tabular}{|c|c|c|c|c|}
\hline

$ b$ &  $ E$   &  $ <r>$ & $ <\frac{1}{r}>$  &     $|\psi(0)|^2$   \\
\hline

1&  0.754098~(0.032)& 2.0231~( 0.079)& 0.7414~(0.025)& 0.129748~(1.05)\\
\hline
2&  0.747750~(0.024)&1.8639~(-0.0061)& 0.6831~(-0.056)& 0.039846~(-0.37)\\
\hline
$\frac{3}{2}$ &0.731223~(0.0007)&1.8668~(0.0046)&0.7197~(-0.0053)&0.062492(-0.013)\\
\hline
$\frac{4}{3}$&0.731726~(0.0014)&1.8910~(0.0084)&0.7301~(0.0091)&0.076171(-0.20)\\
\hline
\end{tabular}
\end{center}
\end{footnotesize}

\vspace{0.5cm}
 
\noindent{\bf III. Trial functions with two, three and  four parameters}

\vspace{0.5cm}

The trial wave function with two or more variational  parameters
may have various forms. The most straightforward one is to
multiply Eq.(7) by a polynomial of $r$, namely
\begin{eqnarray}
\psi(r)=(c_0+c_1~r+c_2~r^2+~\cdot\cdot\cdot~+c_n~r^n)~ e^{-a~r^b},
\end{eqnarray}
where $a$ and ${c_1,c_2,\cdot\cdot\cdot, c_n}$ are  variational parameters,
and $c_0$ can be fixed by the normalization condition. Using the standard
procedure discussed above, we calculate the above mentioned four quantities
in term of the trial wave functions with two, three and four parameters in
the three potential cases and four $b$ value cases. The numerical results
are listed in Tables 2.1$-$2.3, 3.1$-$3.3 and
4.1$-$4.3, respectively.

\vspace{0.5cm}
\centerline{ Table 2.1 }
\vspace{0.3cm}
{\footnotesize
The variational results with the two-parameter trial wave function in the
Cornell potential case. The values listed in parentheses are relative
deviations. The "accurate" results are: $E(1S)=0.257526\; GeV$,
$<r>=1.7073\; GeV^{-1}$,
$<\frac{1}{r}>=0.80848\; GeV$ and $|\psi(0)|^2=0.116054\; GeV^3$.  }

\begin{footnotesize}
\begin{center}
\begin{tabular}{|c|c|c|c|c|}
\hline
  $b$ &  $ E$      & $ <r>$   &     $ <\frac{1}{r}>$ &   $|\psi(0)|^2$\\
\hline

1&0.265416~(0.031)&1.7562~(0.029)&0.8296~(0.026)& 0.158048~( 0.36)\\
\hline
2&0.271382~(0.054)&1.7098~(0.0015)&0.7585~(-0.062)&0.065156~(-0.44)\\
\hline
$\frac{3}{2}$&0.258738~(0.0047)&1.6991~(0.0043)&0.7959~(-0.016)&0.091133(-0.22)\\
\hline
$\frac{4}{3}$&0.257624~(0.0004)&1.7080~(0.0004)&0.8088~(0.0003)&0.108602(-0.064)\\
\hline
\end{tabular}
\end{center}
\end{footnotesize}
 
\vspace{0.3cm}
\centerline{ Table 2.2}
\vspace{0.3cm}

{\footnotesize

The variational results with  the two-parameter trial wave function in
the Martin potential case.
The values listed in parentheses are relative deviations.
The "accurate" solutions are: $E(1S)=7.5605\; GeV$,
$<r>=1.72332\; GeV^{-1}$,
$<\frac{1}{r}>=0.782946\; GeV$, $|\psi(0)|^2=0.0778779\; GeV^3$.    }

\begin{footnotesize}
\begin{center}
\begin{tabular}{|c|c|c|c|c|}
\hline

$ b$ &  $ E$   &  $ <r>$ & $ <\frac{1}{r}>$  &     $|\psi(0)|^2$\\
\hline

1&7.56173~(0.0001)&1.7337~(0.0060)&0.7847~(0.0022)&0.069440~(-0.11)\\
\hline
2&7.56873~(0.0011)&1.7110~(-0.0007)&0.7569~(-0.033)&0.064430~(-0.17)\\
\hline
$\frac{3}{2}$&7.56068~($2*10^{-5}$)&1.7213~(-0.0012)&0.7830~($4*10^{-5}$)
&0.082579~(0.060)\\
\hline
$\frac{4}{3}$&7.56054~($5*10^{-6}$)&1.7222~(-0.0006)&0.7832~(0.0001)&
0.0785065~(-0.0081)\\
\hline
\end{tabular}
\end{center}
\end{footnotesize}
\vspace{0.3cm}
\centerline{ Table 2.3 }
\vspace{0.3cm}

{\footnotesize

The variational results with the two-parameter trial wave function in the
logarithmic potential case. The values listed in parentheses are relative
deviations. The "accurate" solutions are: $E(1S)=0.730733\; GeV$,
$<r>=1.87535\; GeV^{-1}$,
$<\frac{1}{r}>=0.723538\; GeV$,
$|\psi(0)|^2=0.0633063\; GeV^3$.   }

\begin{footnotesize}
\begin{center}
\begin{tabular}{|c|c|c|c|c|}
\hline

$ b$ &  $ E$   &  $ <r>$ & $ <\frac{1}{r}>$  &     $|\psi(0)|^2$   \\
\hline

1&0.731264~(0.0007)&1.88523~(0.0053)&0.7246~(0.0015)&0.058204~(-0.081)\\
\hline
2&0.740704~(0.014)&1.8587~(-0.0089)&0.6968~(-0.037)&0.050334~(-0.21)\\
\hline
$\frac{3}{2}$&0.731022~(0.0004)&1.8707~(-0.0025)&0.7233~(-0.0017)&0.066473(0.050)\\
\hline
$\frac{4}{3}$&0.730798~($9*10^{-5}$)&1.8741~(-0.0007)&0.7236~(0.0001)&0.064889(0.025)\\
\hline
\end{tabular}
\end{center}
\end{footnotesize}

\vspace{0.5cm}
\centerline{ Table 3.1 }
\vspace{0.3cm}
{\footnotesize

The variational results with the three-parameter trial wave function in
the Cornell potential case. The values listed in parentheses are relative
deviations. The "accurate" results are: $E(1S)=0.257526\; GeV$,
$<r>=1.7073\; GeV^{-1}$,
$<\frac{1}{r}>=0.80848\; GeV$ and $|\psi(0)|^2=0.116054\; GeV^3$.}

\begin{footnotesize}
\begin{center}
\begin{tabular}{|c|c|c|c|c|}
\hline
  $b$ &  $ E$      & $ <r>$   &     $ <\frac{1}{r}>$ &   $|\psi(0)|^2$\\
\hline

1&0.257834~(0.0012)&1.7096~(0.0013)&0.8096~(0.0014)&0.108931~(-0.061)\\
\hline
2&0.257908~(0.0015)&1.7063~(-0.0006)&0.8059~(-0.0032)&0.105760~(-0.089)\\
\hline
$\frac{3}{2}$&0.257665~(0.0005)&1.7073~($-4*10^{-7}$)&0.8078~(-0.0008)&
0.105696~(-0.089)\\
\hline
$\frac{4}{3}$&0.257623~(0.0004)&1.7079~(0.0004)&0.8086~(0.0001)&0.108278(-0.067)\\
\hline
\end{tabular}
\end{center}
\end{footnotesize}
 
\vspace{0.3cm}
\centerline{ Table 3.2 }
\vspace{0.3cm}

{\footnotesize

The variational results with the three-parameter trial wave function in
the Martin potential case. The values listed in parentheses are relative
deviations. The "accurate" solutions are: $E(1S)=7.5605\; GeV$,
$<r>=1.72332\; GeV^{-1}$,
$<\frac{1}{r}>=0.782946\; GeV$, $|\psi(0)|^2=0.0778779\; GeV^3$.    }

\begin{footnotesize}
\begin{center}
\begin{tabular}{|c|c|c|c|c|}
\hline

$ b$ &  $ E$   &  $ <r>$ & $ <\frac{1}{r}>$  &     $|\psi(0)|^2$\\
\hline

1&7.56082~($4*10^{-5}$)&1.7264~(0.0008)&0.7843~(0.0017)&0.070081~(-0.10)\\
\hline
2&7.56077~($4*10^{-5}$)&1.7230~(-0.0002)&0.7848~(-0.0023)&0.090033~(0.16)\\
\hline
$\frac{3}{2}$&7.56050~($6*10^{-7}$)&1.7233~($5*10^{-6}$)&0.7829~($-3*10^{-5}$)&
0.077090~(-0.010)\\
\hline
$\frac{4}{3}$&7.56069~($2*10^{-5}$)&1.7236~(-0.0002)&0.7818~(0.0014)&0.081307(-0.044)\\
\hline

\end{tabular}
\end{center}
\end{footnotesize}
\vspace{0.3cm}
\centerline{ Table 3.3  }
\vspace{0.3cm}

{\footnotesize

The variational results with the three-parameter trial wave function in
the logarithmic potential case. The values listed in parentheses are relative
deviations. The "accurate" solutions are: $E(1S)=0.730733\; GeV$,
$<r>=1.87535\; GeV^{-1}$,
$<\frac{1}{r}>=0.723538\; GeV$,
$|\psi(0)|^2=0.0633063\; GeV^3$.   }

\begin{footnotesize}
\begin{center}
\begin{tabular}{|c|c|c|c|c|}
\hline

$ b$ &  $ E$   &  $ <r>$ & $ <\frac{1}{r}>$  &     $|\psi(0)|^2$   \\
\hline

1&0.730759~($4*10^{-5}$)&1.8762~(0.0004)&0.7236~($8*10^{-5}$)&0.065321~(0.032)
\\
\hline
2&0.730964~(0.0003)&1.8757~(0.0002)&0.7245~(-0.0014)&0.072312~(0.14)\\
\hline
$\frac{3}{2}$&0.73088~(0.0002)&1.8745~(-0.0004)&0.7240~(0.0006)&0.0676953(0.069)\\
\hline
$\frac{4}{3}$&0.730796~($9*10^{-5}$)&1.8743~(-0.0006)&0.7238~(0.0004)&
0.0651086 \\
\hline
\end{tabular}
\end{center}
\end{footnotesize}

\vspace{0.5cm}
\centerline{ Table 4.1 }
\vspace{0.3cm}
{\footnotesize

The variational results with the four-parameter trial function in  the
Cornell potential case. The values listed in parentheses are relative
deviations. The "accurate" results are: $E(1S)=0.257526\; GeV$,
$<r>=1.7073\; GeV^{-1}$,
$<\frac{1}{r}>=0.80848\; GeV$ and $|\psi(0)|^2=0.116054\; GeV^3$.  }

\begin{footnotesize}
\begin{center}
\begin{tabular}{|c|c|c|c|c|}
\hline
  $b$ &  $ E$      & $ <r>$   &     $ <\frac{1}{r}>$ &   $|\psi(0)|^2$\\
\hline
 
1&0.257599~(0.0003)&1.7093~(0.0012)&0.8086~(0.0002)&0.114842~(-0.010)\\
\hline
2&0.257667~(0.0005)&1.7021~(0.0003)&0.8096~(-0.0014)&0.111811~(-0.037)\\
\hline
$\frac{3}{2}$&0.257566~(0.0005)&1.7068~(-0.0003)&0.8077~(-0.0010)&0.106024(-0.086)\\
\hline
$\frac{4}{3}$&0.257581~(0.0002)&1.7074~($6*10^{-5}$)&0.8085~($5*10^{-5}$)&
0.110470~(-0.048)\\
\hline
\end{tabular}
\end{center}
\end{footnotesize}
 
\vspace{0.3cm}
\centerline{ Table 4.2  }
\vspace{0.3cm}

{\footnotesize

The variational results with the four-parameter trial wave function in the
Martin potential case. The values listed in parentheses are relative
deviations. The "accurate" solutions are: $E(1S)=7.5605\; GeV$,
$<r>=1.72332\; GeV^{-1}$,
$<\frac{1}{r}>=0.782946\; GeV$, $|\psi(0)|^2=0.0778779\; GeV^3$.    }

\begin{footnotesize}
\begin{center}
\begin{tabular}{|c|c|c|c|c|}
\hline

$ b$ &  $ E$   &  $ <r>$ & $ <\frac{1}{r}>$  &     $|\psi(0)|^2$\\
\hline

1&7.56051~($1*10^{-6}$)&1.7233~($-2*10^{-5}$)&0.7830~($2*10^{-5}$)&0.0791203(0.016)\\
\hline
2&7.56068~($2*10^{-5}$)&1.7192~(-0.0024)&0.7852~(0.0029)&0.0860623~(0.11)\\
\hline
$\frac{3}{2}$&7.56051~($1*10^{-6}$)&1.7226~(-0.0004)&0.7830~($8*10^{-5}$)&
 0.07723~(0.0083)\\
\hline
$\frac{4}{3}$&7.56051~($1*10^{-6}$)&1.7232~($-8*10^{-5}$)&0.7831~(0.0001)&
0.0779022~(0.0003)\\
\hline
\end{tabular}
\end{center}
\end{footnotesize}
\vspace{0.3cm}
\centerline{ Table 4.3 }
\vspace{0.3cm}

{\footnotesize
The variational results with the four-parameter trial wave function in
the logarithmic potential case. The values listed in parentheses are relative
deviations. The "accurate" solutions are: $E(1S)=0.730733\; GeV$,
$<r>=1.87535\; GeV^{-1}$,
$<\frac{1}{r}>=0.723538\; GeV$,
$|\psi(0)|^2=0.0633063\; GeV^3$.   }

\begin{footnotesize}
\begin{center}
\begin{tabular}{|c|c|c|c|c|}
\hline

$ b$ &  $ E$   &  $ <r>$ & $ <\frac{1}{r}>$  &     $|\psi(0)|^2$   \\
\hline

1&0.730737~($6*10^{-6}$)&1.87626~(0.0005)&0.7234~(-0.0001)&0.063531~(-0.0036)\\
\hline
2&0.730913~(0.0002)&1.87499~(-0.0002)&0.72366~(-0.0002)&0.069626~(-0.10)\\
\hline
$\frac{3}{2}$&0.730738~($7*10^{-6}$)&1.8754~($4*10^{-5}$)&0.7236~(0.0001)&
0.064408~(0.017)\\
\hline
$\frac{4}{3}$&0.730732~($1*10^{-6}$)&1.8754~($2*10^{-5}$)&0.7236~(0.0004)&
0.0629413~(-0.0058)\\
\hline
\end{tabular}
\end{center}
\end{footnotesize}

\vspace{0.3cm}

Another possible type of trial wave function with multi-parameters
can be chosen in the form of the superposition of two or more
above mentioned trial wave functions
with different $b$ or $a$ values, respectively. We give an example in
the following. From the resultant $|\psi(0)|^2$ in Tables 1.1$-$1.3, it is
easy to see that in the exponential-type  trial wave function ($b=1$) case,
$|\psi(0)|^2$ values are larger than "accurate"  ones (the deviation
is positive), while in the other three trial wave function cases ($b=2,3/2$
and $4/3$), the corresponding results are less than  "accurate" ones
(negative deviation). Therefore, one can compromise the deviations
by combining two different types of trial wave functions with which
the positive and negative deviations appear, respectively. After
testing various combinations, we find that the results are similar to
those by using three parameters in the last section. In fact, this new
trial wave function has three variational parameters too. In the following,
we list two specific combinations and corresponding results.

\vspace{0.3cm}

The first example is the mixture of a exponential-type wave function and
a Gaussian wave function, i.e.
\begin{eqnarray}
R(r)=c_a e^{-\frac{r}{a}}+ c_o e^{-\frac{\alpha ^2~r^2}{2}}.
\end{eqnarray}

For the Cornell potential, the relative deviation of energy is 0.002 and the
relative deviation of squared WFO is 0.099.

\vspace{0.3cm}

The second example is the mixture of an exponential-type
function and a Gupta's function, i.e.
\begin{eqnarray}
R(r)=c_a e^{-\frac{r}{a}} + c_g~ e^{-a_1 r^{\frac{3}{2}}}.
\end{eqnarray}

For the Cornell potential the relative deviation of energy is 0.00015 and
the relative deviation of squared WFO is 0.025.
     
\vspace{0.3cm}

Apparently, with the same number of variational parameters, these
combined trial wave function can give better description on both energy
and WFO.

\vspace{0.5cm}

\noindent{\bf IV. $2S$ state of $c \bar c$ }

\vspace{0.5cm}

Based on the results of the $1S$ state, it is  easy to discuss the $2S$ state.
We first select a normalized trial wave  function which is orthogonal to
the $1S$ wave function. It can be a single-parameter or
multi-parameter function.
But we find that it is quite difficult to obtain a highly accurate WFO.
In order to make sense, we demonstrate two examples.

\vspace{0.3cm}

The first example is that we take the $1S$ trial wave function as
\begin{eqnarray}
R_{1S}(r)=N~ e^{-a~r^{\frac{4}{3}}},
\end{eqnarray}
where $N$ is the normalization constant and $a$ is determined by the
variational method. Then we choose the $2S$ trial wave function to be
\begin{eqnarray}
R_{2S}(r)=(c_0+c_1~r+c_2~r^2)~e^{-a r^{\frac{4}{3}}},
\end{eqnarray}
where $a$ takes the same value as that in $R_{1S}(r)$. By considering the
orthonormal condition, only one parameter remains free. This parameter can
be fixed by the variational method. The obtained results
show that the relative deviation of energy is 0.004, while the relative
deviation of the squared WFO of the 2S state is 0.25.

\vspace{0.3cm}

Moreover, if we take
\begin{eqnarray}
R_{2S}(r)=(c_0+c_1 r+c_2 r^2 +c_3 r^3 )~e^{-a r^{\frac{4}{3}}},
\end{eqnarray}
there are two variational parameters in the trial wave function.
The resultant relative deviations of energy  and  squared WFO
turn to be  0.0009 and 0.11, respectively.

\vspace{0.3cm}

Similar to the $1S$ state trial wave function, if we take a trial wave
function with four parameters, the resultant relative deviation of the
squared WFO of the $2S$ state is less than 0.05.

\vspace{0.5cm}

\noindent{\bf V. Relationship between WFO and reduced mass}

\vspace{0.5cm}

There were lots of discussions concerning the relation between
WFO and the reduced mass in the heavy quarkonium system [1,2].
Because the variational method can give the analytical expression of the
wave function, consequently the exact value of WFO, the approximate relation
between WFO and the reduced mass can be obtained. For example,
if the trial wave function with parameter $a$ for the $1S$
state, Eq.(7), is chosen, the squared WFO can be written as
\begin{eqnarray}
|\psi(0)|^2=N^2=\frac{b~(2a)^{\frac{3}{b}}}{4 \pi \Gamma(\frac{3}{b})}.
\end{eqnarray}
In the Martin potential case, substituting (14) into (22), we obtain
\begin{eqnarray}
|\psi(0)|^2=
\frac{b}{4 \pi \Gamma(\frac{3}{b})}
       [\frac{0.4 k\Gamma(\frac{3.1}{b})}
          {b^2\Gamma(2+\frac{1}{b})}~\mu]^{\frac{3}{2.1}}.
\end{eqnarray}
This relation is similar to that given by the simple scaling
arguments for the power-law potential, namely
\begin{eqnarray}
|\psi(0)|^2\sim \mu^{\frac{3}{2+\nu}}.
\end{eqnarray}
When $\nu=0.1$, Eqs.(23) and (24) coincide with each other.
It clearly shows that although the variational method is an approximation,
it retains the main characteristics of the solution. This observation
encourages us to apply the variational method to the estimation of WFO.

\vspace{0.3cm}

As shown in Table 1.1, when $b=\frac{3}{2}$, the squared WFO of the $1S$
state of $c \bar c$ has the least deviation(0.027). By taking
$m_b=5.174 GeV$ \cite{mar}, one obtains 0.361475 for the squared WFO of
the  $1S$ state of $b \bar b$. In the $B_c$ meson
case, the calculated reduced mass is 1.3336 $GeV$, and the squared WFO
of the $1S$ state is 0.140547. The resultant relative deviations of WFO
for both $b\bar b$ and $B_c$  are 0.027 which is the same as that
for $c\bar c$.

\vspace{0.3cm}

In the logarithmic potential case,  the similar treatment leads to
\begin{eqnarray}
|\psi(0)|^2=
\frac{b}{4 \pi \Gamma(\frac{3}{b})}
       [\frac{4 k\Gamma(\frac{3}{b})}
          {b^2\Gamma(2+\frac{1}{b})}~\mu]^{\frac{3}{2}}.
\end{eqnarray}
It is easy to see that the relation between WFO and the reduced mass $\mu$
is consistent with that obtained from the scaling arguments
for the power-law potential with $\nu=0$.

\vspace{0.3cm}

Again in the single-parameter trial wave function with $b=\frac{3}{2}$ case,
one finds the least deviation. By taking
$m_c=1.5 GeV$ and $m_b=4.906 GeV$ \cite{qr1}, the squared WFOs of the $1S$
states of $J/\psi$, $\Upsilon$ and $B_c$ are
0.062492, 0.36964 and 0.118462, and the corresponding relative
deviations are about 0.013, respectively. This is a quite satisfactory
and encouraging observation.

\vspace{0.3cm}

In the Cornell potential case,  by employing the single-parameter
trial wave function with $a$  given in Eq.(12), one can carry out a
similar discussion. The obtained relation between the squared WFO
and the reduced mass is much more complicated than that in Eqs.(23) and (25).
We omit the detailed  expression in the text but give following two
interesting results, which are computed by taking $b=\frac{4}{3}$ and
the relative deviation within $10 \%$.

\vspace{0.3cm}

(1). The curve of  $|\psi(0)|^2$ as the  function of the reduced mass $\mu$
has the similar behavior as those in the former two cases. When
$\mu=0.2-3.0GeV$, the function $f(\mu)\sim ~\mu^{2.2}$ can  fit the
curve very well. This corresponds to  $\nu=-0.65$ in Eq.(24).

\vspace{0.3cm}

(2). If we take $m_c=1.84~GeV$ and $m_b=5.17~GeV $ \cite{egkl},
the conjecture of Eq. (8) in Ref.\cite{cd}

$$ |\psi_{b \bar c}(0)|^2 \simeq
  |\psi_{c \bar c}(0)|^{1.3}~|\psi_{b\bar b}(0)|^{0.7} $$
holds within $2.5 \%$.

\vspace{0.5cm}

\noindent {\bf VI. Conclusion and discussion}

\vspace{0.5cm}

In this paper, we carefully studied the variational method, especially
in determining the binding energies, wave functions at
the origin, average radii and etc. of the quarkonium.
Retaining generality as much as possible,
we employ several "successful" potential models to analyze. By
comparing the numerical results obtained in terms of the variational method
with those by solving Schr\"{o}dinger equation numerically, we
present the relative deviations by employing different trial wave
functions with single- or multi-parameters in various potential models.

\vspace{0.3cm}

In the variational method with fewer parameters, even the calculated binding
energies and some transition matrix elements are accurate enough, it is not
easy to obtain a very accurate WFO. Because of the phenomenological
requirements, a certain accuracy is requested. Thus, one of our aims
is to seek for a possible and simpler way to solve this problem.

\vspace{0.3cm}

We first study the single-parameter trial wave function case. The results
shown in Tables 1.1-1.3 indicate that for the Cornell potential, the trial
wave function with $b=\frac{4}{3}$, i.e. $\psi(r)=N~e^{-a~r^{\frac{4}{3}}}$,
can give the least relative deviation of squared WFO. The value of the
deviation is about 0.11 which is not accurate enough, although the relative
deviation of energy already reaches $10^{-3}$. For the Martin and
the logarithmic potentials, the situations are better. When
$b=\frac{3}{2}$, one obtains the least values of 0.027 and 0.013 for the
relative deviations of squared WFO, respectively, while the corresponding
relative deviations of energy reach $10^{-4}$ and $10^{-5}$, respectively.

\vspace{0.3cm}

The accuracy of variational results can be improved when the number of
the variational parameters are increased. This is true in general.
If one adopts more than four variational parameters, he may expect very high
accuracy for energy, but not always for WFO. It is much
more difficult to further improve the accuracy of WFO than that of energy.
The resultant accuracy of WFO seriously depends on the choice of the
trial wave function. Sometimes as the number of parameters increases,
the accuracy of WFO deteriorates, while the accuracy of energy is
much more improved. Of course, in this case,
the variational calculation becomes much more tedious and difficult.

\vspace{0.3cm}

The trial wave function with a single variational parameter is
most convenient for use. If the accuracy of $10\%$ for WFO in
the Cornell potential case is tolerable,  $\psi(r)=N~e^{-a~r^{\frac{4}{3}}}$
would be the best choice for the $1S$ state trial wave function. For the
Martin and logarithmic potentials, $\psi(r)=N~e^{-a~r^{\frac{3}{2}}}$ is
the most appropriate trial wave function for the $1S$ state, and the
corresponding WFOs have quite satisfactory accuracies. In particular, these
forms can give very simple and reasonable relations between WFO and the
reduced mass, which agree with those deduced from the general arguments for
power-law potentials. However, the requested accuracy of squared WFO
in the system concerned is generally lower than $2-3\%$, and the trial wave
functions with a single parameter cannot provide this accuracy. For instance,
one indicated in Ref.\cite{am} that the estimated decay constant in
Ref.\cite{is}, where a single-parameter harmonic oscillator trial wave
function was employed in solving the Schr\"{o}dinger
equation in the Cornell potential case, may not be reliable, although
the computed transitions and energy are quite reasonable.

\vspace{0.3cm}

Here we would like to point out that
to solve the Hamiltonian (Eq.(77) in Ref.\cite {ty})
\begin{eqnarray}
H^{(1)}=-\frac{1}{m} \Delta-\frac {C_F \widetilde{\alpha}_s}{r}-
    \frac{C_F~\beta_0~\alpha_s^2}{2 \pi}~\frac{ln~r \mu}{r},
\end{eqnarray}
the exponential-type trial wave function with a single parameter (Eq.(79) in
\cite{ty})
\begin{eqnarray}
f_b(r)=\frac{2}{b^{\frac{3}{2}}}~e^{-\frac{r}{b}}
\end{eqnarray}
was employed in the variational framework. The paper reported that
because the resultant energy agreed with the "accurate" value up to
an order of $O(\alpha_s^4)$, Eq.(26) could be an appropriate trial wave
function. Then, the resultant WFO would be accurate enough to serve
their main goal. However, our estimation shows that by using various
parameters given in Eq.(26) (take $\mu=m$ in Eq.(27)), the relative
deviations of squared WFO are about 0.40 and 0.18 for $c \bar c$
and $b \bar b$, respectively. These results would
lead to improper  theoretical predictions for the quantities which are
closely related to WFO.

\vspace{0.3cm}

As the conclusion we can draw, for the binding energy, most of
trial wave functions, even with a single-parameter, can result a solution
with a satisfactory accuracy. However, This is not always true for WFO.
To obtain a reliable value of WFO, one has to adopt not only an appropriate
trial wave function form for a specific potential, but also
the proper number of variational parameters in the function.
Our finding indicates that for a specific potential form and a not very
higher accuracy of WFO, one can always find out a relatively simpler and more
reliable trial wave function with an appropriate number of parameters.
However, the form of the trial wave function strongly depends on the
potential. In general, there is no  universal rule to determine the form
and the number of the parameters of the trial wave function.

\vspace{0.3cm}

On the other hand, our study shows that if one chooses Eq.(16) as the trial
wave function,  when the potential is flatter at the large $r$, the
higher power terms, namely the higher configuration
mixing, should be considered so that the higher accuracy of WFO can be
reached. Moreover, the results by using Eqs.(17) and (18) indicate that
if several components of the trial wave function can compromise
the descriptions of the asymptotic behaviors of the potential at the short-
and long-ranges, respectively,  the trial wave function would have the
simplest form and can provide higher accuracies for both binding energy
and WFO. Namely, the trial wave function is more efficient. Usually, when
a four-parameter trial wave function is chosen, in the commonly used
potential cases, the accuracies of the energy and WFO can reach $10^{-4}\sim
10^{-6}$ and $10^{-2}\sim 10^{-3}$, respectively.

\vspace{0.3cm}

As mentioned in the introduction, the variational method should further
be studied. Now, if the potential is not very singular, namely
would not cause the divergence in solving Schr\"{o}dinger equation, we
find the way to construct an efficient trial wave function to get the more
accurate binding energy and WFO. Then the next step is to study that if
the potential is very singular, say more singular than $1/r^{3}$, how the
bound state problem can be solved and the real physics can be obtained
by using the variational method, and whether the obtained result in this
method corresponds to reality. These discussion is shown in our next paper
\cite{dls}.

\vspace{1cm}

\noindent {\bf Acknowledgments}

\vspace{0.5cm}

One of the authors (Y. B. Ding) would like to thank Prof. G. Prosperi for his
hospitality during his stay in the Department of Physics, University of
Milan.

\end{document}